# Superdielectric Materials Composed of NaCl, $H_2O$ and Porous Alumina

By Jonathan Phillips
Prof. Physics
Naval Postgraduate School
Monterey, CA 93950

jphillip@NPS.edu

ABSTRACT

To test a theory of the recently discovered phenomenon of superdielectric behavior the dielectric constants of several 'pastes' composed of porous alumina powders filled to the point of incipient wetness with water containing dissolved sodium chloride, were measured. The dielectric constants of some of the pastes were greater than $10^{10}$, higher than that of any material ever reported. These results are consistent with this recently postulated model of superdielectric behavior in porous, non-conductive materials saturated with ion-containing liquids: Upon the application of an electric field ions dissolved in the saturating liquid contained in the pores will travel to the ends of pore filling liquid droplets creating giant dipoles. The fields of these giant dipoles oppose the applied field, reducing the net field created per unit of charge on the capacitor plates, effectively increasing charge/voltage ratio, hence capacitance. Other observations reported herein include; i) the impact of ion concentration on dielectric values, ii) a maximum voltage similar to that associated with the electrical breakdown of water, iii) the loss of capacitance upon drying, and iv) the recovery of capacitance upon the addition of water to a dry super dielectric material. All observations are consistent with the earlier proposed model of the super dielectric phenomenon.

## INTRODUCTION

Recently a material with an extraordinary dielectric constant, greater than $10^8$, was described (1). This material, dubbed a 'super dielectric', was composed of a high surface area alumina powder saturated with a solution of boric acid (pH ~5) to the point of 'incipient wetness'. Capacitors generated with this material, New Paradigm Super (NPS) capacitors, behaved as electrostatic capacitors below about 0.8 V. Extrapolating the data to 5 micron layers, a thickness regularly attained with standard 'ceramic' capacitors suggested that this material could be used in NPS capacitors easily capable of storing energy at about 50 J/cm$^3$. Moreover, a theory was advanced that suggested that a family of superdielectric materials exists, and that some members of this family of materials could achieve far higher energy densities. The theory: Any porous, non-electrically conductive, solid saturated with a liquid containing dissolved ions, is potentially a super dielectric material, because of the fundamental mechanism of super dielectric behavior. Specifically, in an electric field, ions in the liquid drops move to form dipoles that 'oppose' the applied field. This increases the charge on the electrodes required to produce a given net voltage, hence increasing the q/V ratio, that is, the capacitance. The maximum voltage such materials can sustain is limited by the breakdown voltage of the liquid phase.

A simple illustration of the practical implications of the model: If the same alumina employed in the first study, were to be saturated with an electrolyte with a higher discharge voltage than water, for example 2.5 volts, but containing the same amount of dissolved boric acid, an NPS type capacitor of 5 microns thickness would be able to store >600 J/cm$^3$. Clearly, the model should be tested as it suggests the original report represents only one example of a broad class of materials. The arbitrary 'first example' of the earlier study is unlikely to represent an optimum formulation. Hence, testing aspects of the model may lead to the discovery of materials with even higher dielectric constants.

The work described herein was deliberately limited to permit a focus on the potential value of the NPS capacitor for electric energy storage. Thus, the study was limited to measuring capacitance at ~0 Hz. The study does not address the issue of using an NPS capacitor as a circuit element; although such studies are clearly a component required to fully understand super dielectric materials. Efforts were directed to test elements of the model as applied to the 'zero hertz' case. In particular, studies were designed to test the postulate that any ionic solution added to the same alumina used in the first study could produce super dielectric behavior. To do this, only the source of the ions was changed, and all other parameters, particularly the alumina used, the use of water as the electrolyte, the ratio of water/alumina used to form the paste, the physical dimensions of the capacitors, etc, were (nearly) unchanged. The only significant change was the use of NaCl rather than boric acid as the source of dissolved ions in the water.

The results were completely consistent with the earlier model; super dielectric behavior was observed. Moreover, the dielectric constants observed for two of the pastes ($\sim 10^{10}$), were a full order of magnitude higher than reported in the first paper on super dielectrics.

EXPERIMENTAL

Dielectric Fabrication: The materials employed to create the specific dielectric employed in this study, alumina/NaCl solution super dielectric material (Salt-SDM), were high surface area aluminum oxide powder (Alfa Aesar, γ-phase, 99.97%, 3 micron APS Powder, S.A. 80-120 $m^2/g$, CAS 1344-28-1), NaCl powder (Sigma-Aldrich 10 mesh anhydrous beads, 99.999%), and distilled deionized water. These constituents were mixed by hand in three ratios. In all cases this alumina:$H_2O$ ratio was the same 1 g alumina:1.1 mL $H_2O$. Three different NaCl ratios were employed: Low Salt: 0.01 gm salt/1 gm alumina; Medium salt: 0.1 gm salt/1 gm alumina, and High salt: 0.3 g salt/1 gm alumina. In each case the mixing process was as follows. First, water was added to a plastic cup, next salt was added. These mixes were agitated by shaking, not stirring, until all the salt was dissolved in the water. Finally, alumina, in the proper ratio, was gradually added. This created a spreadable paste with little or no 'free' water (incipient wetness). It is interesting to note that salt and water are pH neutral.

As pore structure is a significant component of the proposed model, the surface area and pore structure were determined from BET nitrogen isotherms collected at 77K and analyzed using a Quantachrome NOVA 4200e. Two samples were independently measured and both yielded results within 5% for all parameters; specifically a surface area of 39 +/-1 $m^2/g$, a total pore volume of 0.45 $cm^3/g$ and an average pore radius of 245 +/- 3 Å.

The dielectric paste was spread evenly on a 5 cm diameter disc of GTA grade Grafoil (0.76 mm thick, >99.99% carbon). As described elsewhere (2, 3) Grafoil is a commercially available high purity carbon material (available in sheets or rolls) made by compressing naturally occurring graphite flakes with a surface are on the order of 20 $m^2/g$. In the final step a second sheet of Grafoil was placed on top, then mechanically pressed to create a near constant thickness as determined by measurements made at multiple positions using a hand held micrometer. This step completed the construction of an NPS Capacitor. The 'effective thickness' of the dielectric, required to compute the dielectric constant, used in all computations herein was based on subtracting the Grafoil sheets thickness from the measured gross thickness of the capacitor. The measured thicknesses of the dielectric layers are as follows: low salt 0.64 +/- 0.08 mm, medium salt 0.50 +/- 0.06 mm, high Salt 0.46 +/- 0.02 mm. The errors are the spread in the eight thickness values measured. These 'errors' probably accurately reflect real, but small, variations in the NPS Capacitor thickness.

Once constructed the capacitors were placed in an electrically insulating plastic jig with bottom and top cylindrical aluminum electrodes of 5 cm diameter and 5 mm thickness. A 250 g weight was placed on top in all cases. These capacitors were then placed in simple circuits (Figure 1) for measurements of charge and discharge. It is important to note that in all cases in this study charging and discharging was through a 7.5 kOhm resistor for the low and medium salt capacitors, and 20 kOhm for the high salt capacitor. The use of a simple circuit, rather than a commercial meter, to measure capacitance and subsequently dielectric constant, was dictated by the goal of this program: measuring the ability of the NPS capacitor to store electrical energy. Indeed, commercial meters intended to measure capacitance employ an algorithm. The actual data is collected at very low voltages (<<1 Volt) only, and then 'deconstructed' to yield dielectric constant, although the better meters yield this deconstructed dielectric value as a function of frequency. In order to evaluate the capacity for electrical energy storage, maximum operating voltage, and ~0 Hz data, is needed. These values are not available from meters.

The primary test platform was a National Instruments ELVIS II electronics prototyping board implemented with LabView 2011 software. An additional multimeter, Agilent U1252A, was regularly used for independent parameter verification. It is further notable that the capacitance of several types of commercial capacitors were measured using the above described instruments and protocol, and in every case the measured value and the listed value were within 30%.

## RESULTS

The charging and discharging behavior of three NPS capacitors was studied. As the effect of ion concentration on capacitive behavior was the target of the investigation, the salt concentration was varied by a factor of 30 over these three NPS capacitors. There were also small differences in the thickness of the dielectric layer (less than a factor of 1.5), an inevitable consequence of the imprecise nature of the hand construction. However, it was established in the first paper on superdielectrics that the dielectric constant of super dielectric materials is nearly independent of layer thickness (1).

Low Salt NPS Capacitor- Repeated charge/discharge cycles for this capacitor are shown in Figure 2. Although the applied charging voltage was 4V, the NPS Capacitors never reached more than about 1.8 volts. From Figure 2 it is readily apparent that this capacitor, which is exemplary of all three studied, discharged in stages; initially very rapidly, down to approximately 1.1 Volt, then much more slowly. As discussed in detail below, the slow discharge region, below ~1 Volt, also can be divided into two different discharge regimes.

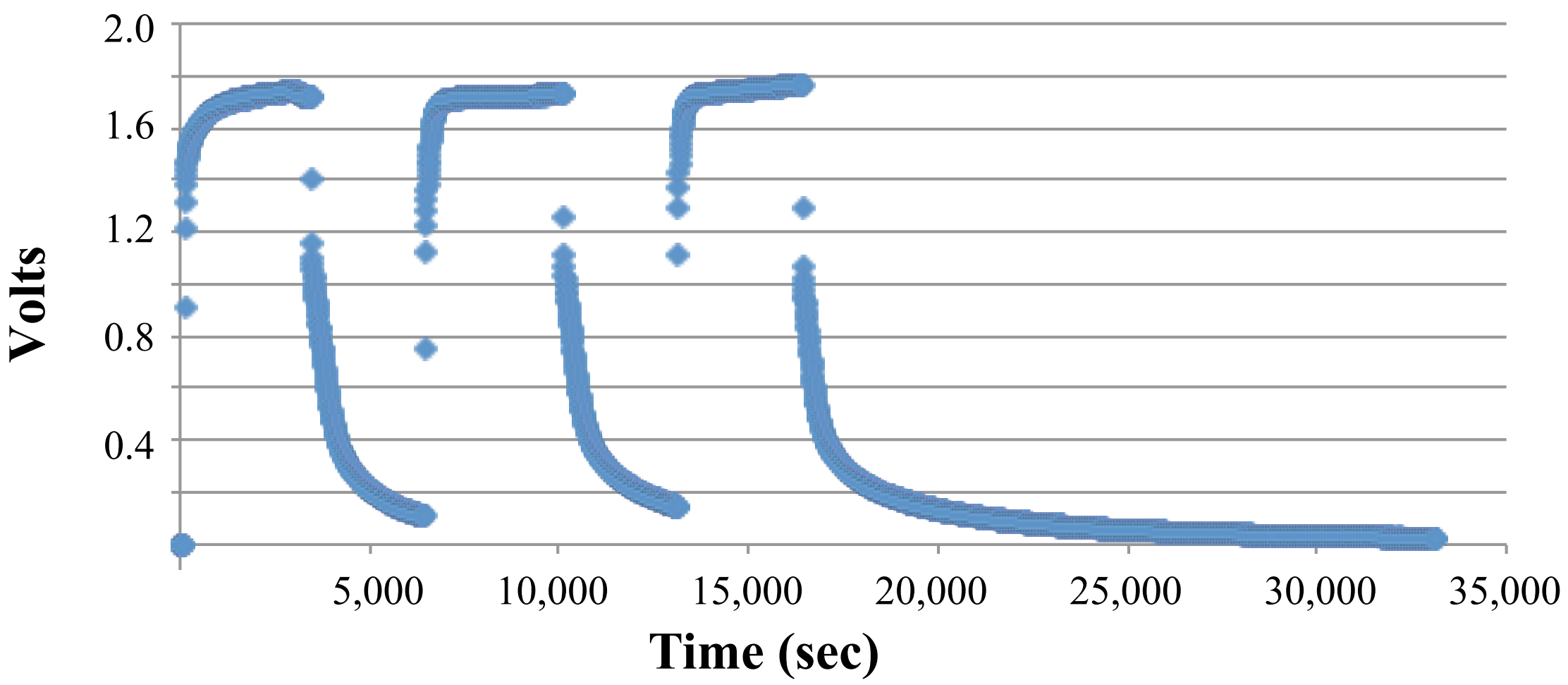

**Figure 2**- *Low Salt Cycles*. Three cycles of charge and discharge are shown. The low salt NPS Capacitor never charges above 1.8 Volts, but clearly can discharge to zero and can be repeatedly recharged.

To explore the discharge behavior in a quantitative fashion, the discharge voltage/time data from the three cycles shown in Figure 2 were plotted, below 1.1 volts in a semi-log form (Figure 3). For a constant capacitance these plots are linear, and show two linear regions of different slope with an 'elbow' between. In sum, the data can best be modeled as showing three regions of capacitance. The first region is between the highest voltages reached during charging, ~1.8 V, and about 1.1 volts. In this region the capacitance is low, and no effort was made to determine the actual value. The data is not plotted in Figure 2. The second region is for voltages between ~1.1 Volts and ~350 mV. In this region the capacitance is very high and consistent for all three cycles. The third region is voltages below ~300 mV. In this region the capacitance is 'off scale', relative to commercial ceramic capacitors of the same size. Actual capacitance values are given in Table I.

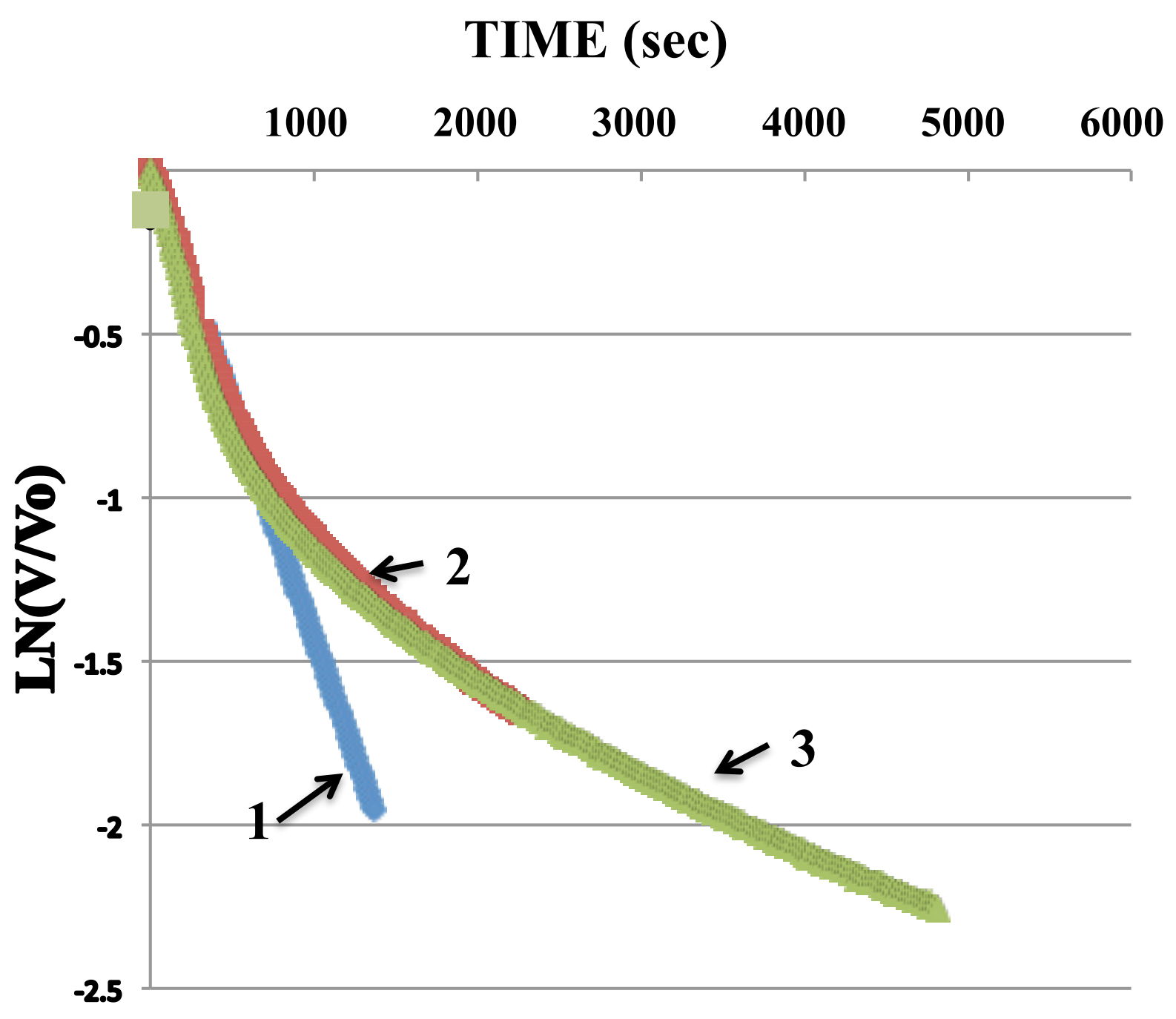

**Figure 3**- *Two Regions of Capacitance*- The numbers refer to the discharge cycles in Figure 2, and the V0 value in all cases was 1.1 V. Clearly for cycles 2 and 3 there is a sharp change in capacitance at about 350 mV. (The first cycle has a less 'sharp' elbow that is hidden in this figure.)

| SALT LEVEL | CYCLE | VOLTAGE | CAPACITANCE (F) | DIELECTRIC CONSTANT | Max Error % Capacitance and Dielectric |
|---|---|---|---|---|---|
| Low | 1st discharge, high volts | 1-0.3 | 0.086 | $3.1\ 10^{9}$ | +/- 20% |
| Low | 1st discharge, low volts | <0.3 | 0.093 | $3.4\ 10^{9}$ | +/- 20% |
| Low | 2nd discharge, high volts | 1-0.35 | 0.11 | $3.9\ 10^{9}$ | +/- 20% |
| Low | 2nd discharge, low volts | <0.30 | 0.306 | $1.1\ 10^{10}$ | +/- 20% |
| Low | 3rd discharge, high volts | 1- 0.35 | 0.09 | $3.3\ 10^{9}$ | +/-20% |
| Low | 3rd discharge, low volts | <0.30 | 0.45 | $1.6\ 10^{10}$ | +/-20% |
| Medium | 1st discharge, High Volts | 0.8-0.3 | 0.98 | $2.7\ 10^{10}$ | +/- 25% |
| Medium | 1st discharge, Low Volts | <0.3 | 2.25 | $6.3\ 10^{10}$ | +/-25% |
| Medium | 2nd discharge High Volts | 0.9-0.3 | 1.05 | $3.1\ 10^{10}$ | +/-25% |
| Medium | 2nd discharge Low Volts | <0.3 | 2.25 | $6.3\ 10^{10}$ | +/-25% |
| Medium | After Dry and Re-wet | 0.9- 0.3 | 0.80 | $2.2\ 10^{10}$ | +/-30% |
| High | 1st discharge, High Volts | 1.0-0.3 | 0.17 | $4.4\ 10^{9}$ | +/-20% |
| High | 1st discharge, Low Volts | <0.3 | 1.49 | $3.9\ 10^{10}$ | +/-20% |
| High | 2nd discharge, High Volts | 1.0-0.3 | 0.24 | $6.2\ 10^{9}$ | +/-20% |
| High | 2nd discharge, Low Volts | <0.3 | 4.7 | $1.2\ 10^{11}$ | +/-20% |
| High | 3rd discharge, High Volts | 1.1-0.3 | 0.19 | $4.9\ 10^{9}$ | +/-30% |
| High | 3rd discharge, Low Volts | <0.3 | 2.49 | $6.5\ 10^{10}$ | +/-30% |

Medium Salt- The discharge behavior of the medium salt and low salt NPS Capacitors were qualitatively similar. Both discharged rapidly above a particular voltage, and both showed an an 'elbow' in the discharge curve at a low voltage that corresponded to a change in capacitance (Figure 4). However; the quantitative values associated with these features were different. For the medium salt the high capacitance/slow discharge point in the medium salt was lower, about 0.9 volts, rather than 1.1 volts. Also, the dielectric constants for the medium salt were consistently higher. During the high volt leg the low salt dielectric constant was fairly consistent over three discharge curves, equal 3.5 +/- 0.4 $10^9$, whereas the medium salt dielectric constant was never less than 2.7 $10^{10}$, nearly 8X higher. The dielectric constant of the low voltage leg of the medium salt was at least 4X higher than that observed at low salt

In order to test the 'liquid dipole' aspect of the theory the medium salt NPS capacitor the medium salt capacitor was allowed to dry, its capacitance measured, and then the paste was 're-wetted' and its capacitance measured again. According to the theory, an NPS capacitor should have virtually no capacitance after drying, but capacitance should be restored with the addition of water. All observations were consistent with these expectations of theory as described in more detail below

To test the 'dry state', the medium salt NPS capacitor was allowed to sit in the room for 15 days. Indeed, as shown in Figure 5, at the end of this period of time the 'paste' appeared dry and cracked. In this state the medium salt NPS capacitor had a measured resistance of ~5 MOhms, and no measurable capacitance. That is, after charging for more than an hour, the capacitor discharged to ~1 mV in less than 10 seconds. The measured near zero capacitance is consistent with the behavior anticipated by theory: In the absence of water in the pores, there can be no 'giant dipoles', hence no super dielectric behavior.

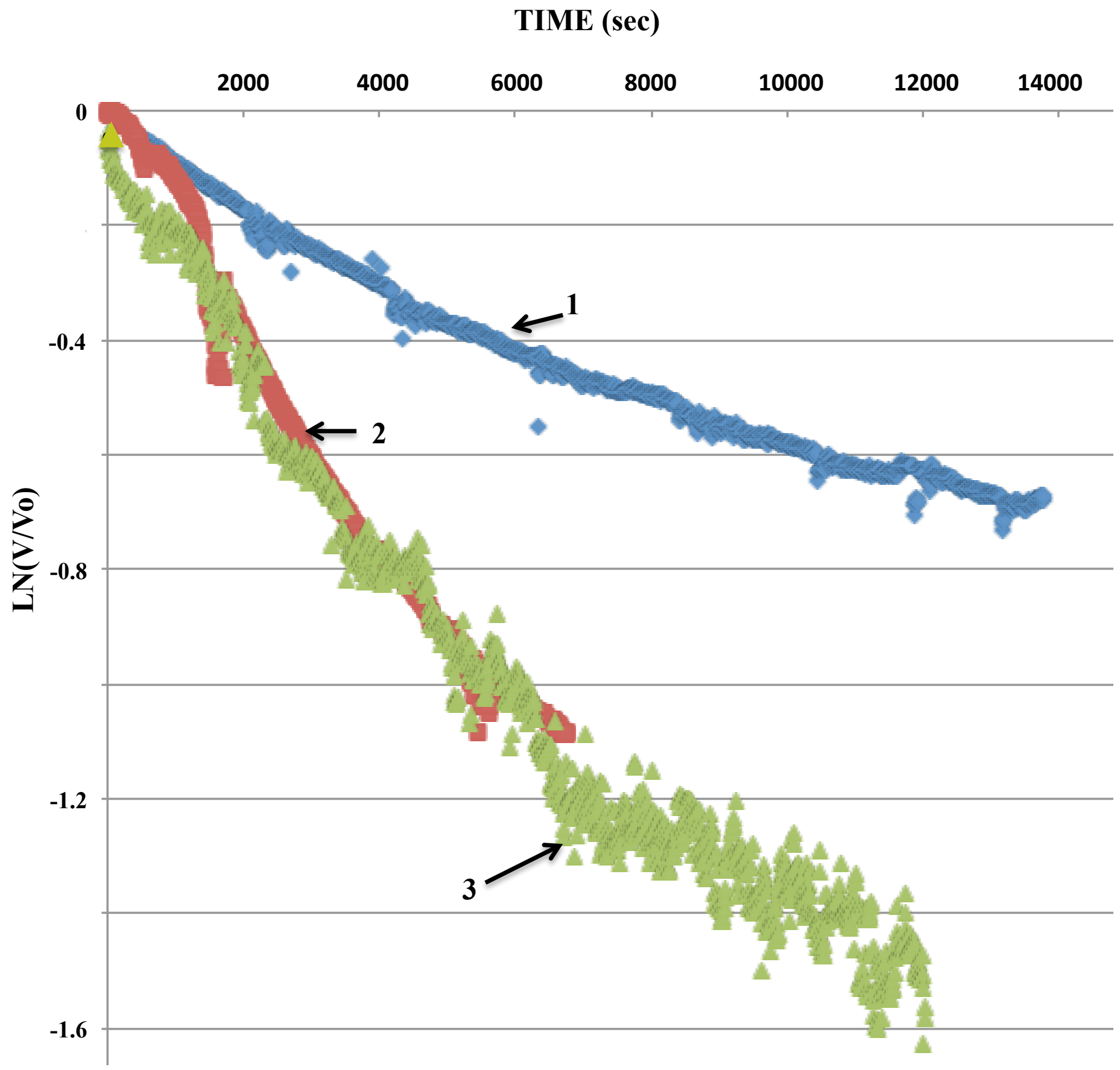

**Figure 4**: *Medium Salt NPS Capacitor Discharge*. 1- At Low Voltage (<`0.3V) the material has a very high dielectric constant. 2- At High Voltage (0.3<V<0.9) the dielectric constant is about 50% lower. 3- After drying this capacitor had approximately zero capacitance, but the original high voltage capacitance was mostly restored, as shown, by the addition of water (see text).

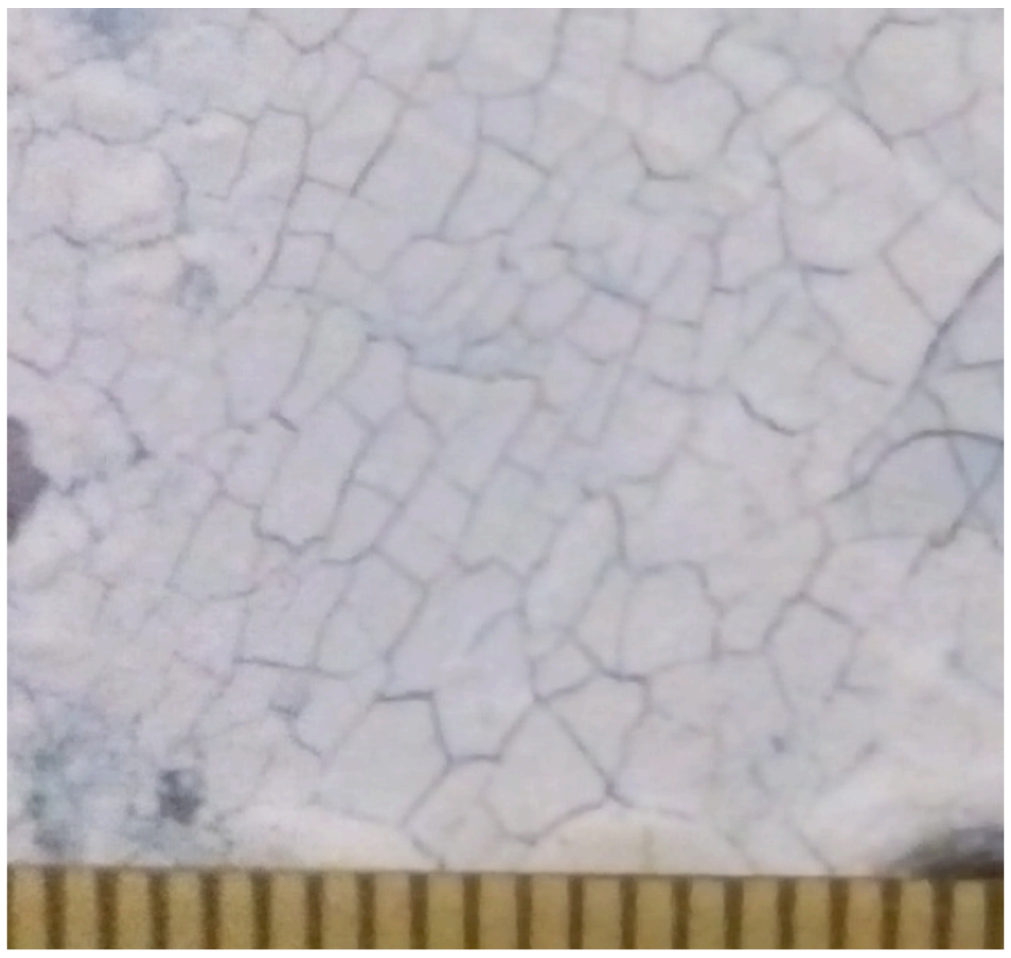

**Figure 5**- *Dried Paste*- Fifteen days after the medium salt NPS capacitor was created there was no measurable capacitance and, as shown, the paste is dried and cracked. Each line on the scale is 1/32”.

As noted above, the following is a logical correlation to the giant dipole theory: A dried and ‘dead’ NPS capacitor can be restored to initial performance by the addition of water. Indeed, as the salt content is not modified by the drying process, the addition of water to the pores should re-dissolve the salt and permit giant dipoles to form. Concomitantly super dielectric behavior should be restored. Hence, water was added to the dried paste of the medium salt NPS capacitor by evenly spreading water, roughly equal to the amount initially present in the dielectric based on an initial measure of the weight of the paste employed in creating the NPS capacitor, on one side of that Grafoil sheet which was removed to permit the inspection (Figure 5). This sheet, water side down, was then pressed back onto the dielectric, reforming the NPS capacitor. The result of this restoration protocol can be seen in Figure 4 (line 3): It nearly restores the initial dielectric constant. There is one difference between the initial behavior of the medium salt NPS capacitor and the restored medium salt NPS capacitor: below about 250 mV there is an ‘elbow’ but the data becomes very noisy. Hence, no low voltage capacitance value is provided. The origin of this behavior is not clear.

High Salt- The qualitative discharge characteristics of this NPS Capacitor were similar to the other two. First it can be, and was, repeatedly cycled, as per Figure 2. Second, there was a region of low capacitance from about 1.8 V to 1.0 volts. Third, the slow discharge region could be divided into two sections, a section of high capacitance between about 1 V and 300 mV and a section of extremely high capacitance below 300 mV.

The quantitative behavior was unique. At ‘high voltages’, roughly from 1 V to 300 mV, this capacitor showed performance similar to that observed for the low salt case.

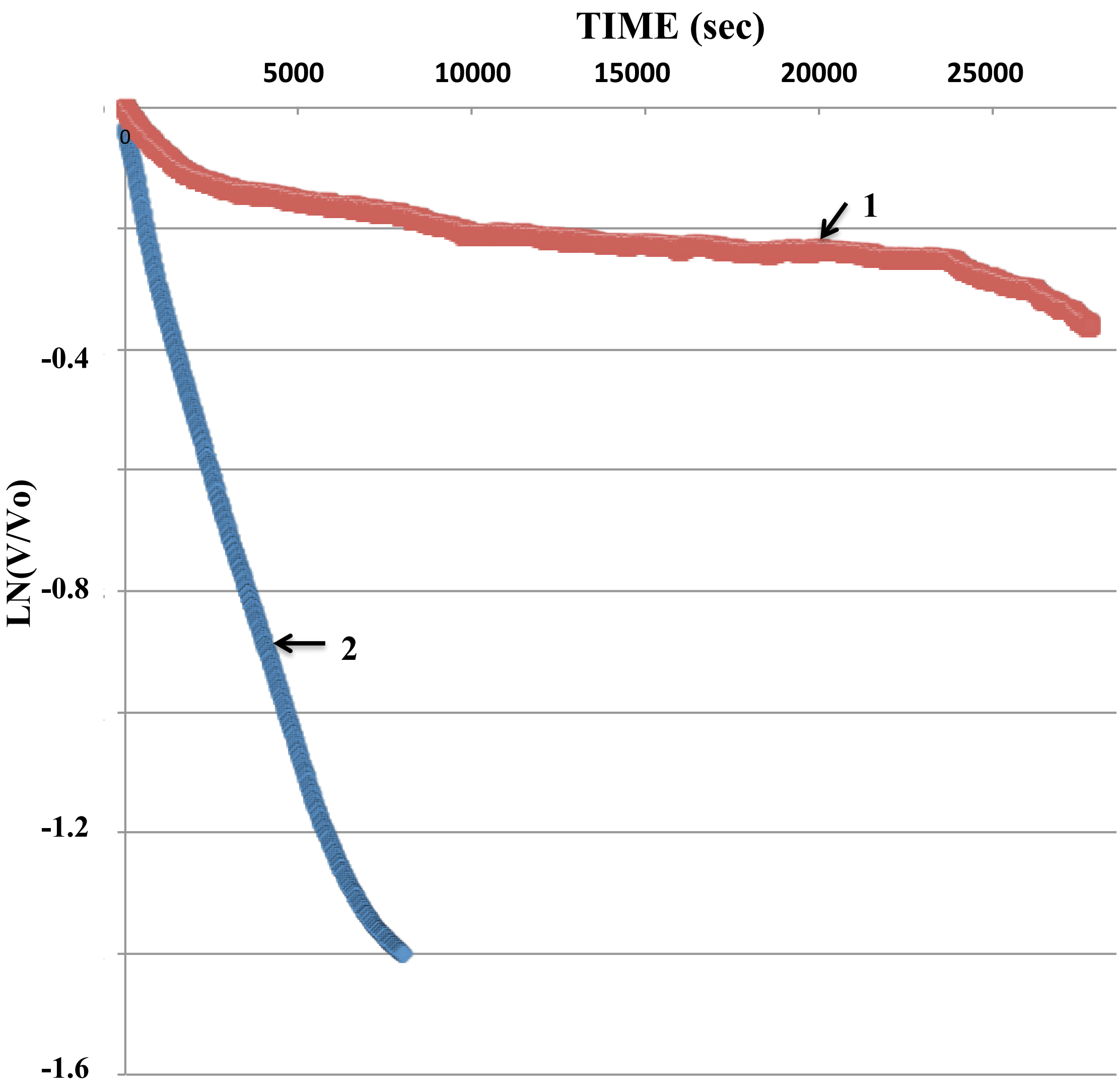

**Figure 6**- *High Salt NPS Capacitor.* The second discharge cycle behavior is graphed as two regions of dielectric performance. Line 1- The capacitance is extremely high ($>1*10^{11}$) below 300 mV. Line 2- The capacitance between 1.0 V and 300 mV ($>6*10^{9}$) is similar to that observed for low salt NPS capacitors 300 mV.

At best over this voltage range, the dielectric constant was no more than 2X that observed for the low salt sample, and clearly far less than that observed for the medium salt sample. In sum, the measured dielectric constant over this voltage region is not a linear function of salt concentration. Below 300 mV there is a very sharp change in the dielectric value (Figure 6). In this voltage range the measured dielectric is similar to that observed for the medium salt, in fact on average 15% higher. For this low voltage range there apparently is a positive correlation between dielectric constant and salt loading, one that perhaps reaches a limit asymptotically with the concentration of salt.

## DISCUSSION

There are four main points of this research. One, it confirms that superdielectric behavior at ~0 Hz is repeatable. Two, various parameters can be manipulated to increase the dielectric constant. Three, the results are consistent with the hypothesis that super dielectric behavior arises from the formation of dipoles in ion containing liquid drops in pores, and that consequently super dielectrics are a broad family of materials. Four, it is reasonable to extrapolate the results to suggest the potential energy density of NPS capacitors can be enormous, possibly surpassing the current generation of Li ion batteries.

In the first paper on super dielectrics, an aqueous boric acid solution was used to create the alumina paste (1). It was postulated that there was nothing essential about boric acid and that any solution containing dissolved ions would perform as well. In this paper aqueous solutions of NaCl were used to saturate the alumina, and the resulting dielectric constants were about an order of magnitude higher than that of those produced with boric acid. These results are not regarded as simply empirical evidence that NaCl will produce a better dielectric constant than boric acid. More broadly the results suggest that many parameters can be manipulated to optimize dielectric constant. For example in the present work it was established that up to a certain concentration ‘more salt is better’. In fact, the results suggest there may be an optimum salt concentration. The ‘medium’ salt capacitor had a higher energy density than the ‘high’ salt sample This suggests there may be an optimum pore size, an optimum porous refractory, an optimum electrolyte, an optimum salt, etc.

The NPS capacitor is not the only example of an inhomogeneous material with a high dielectric constant. Although virtually all commercial dielectrics are homogeneous materials, in particular Barium Titanate, and work continues on further developing this material (4-7), there are two classes of non-homogenous materials that reportedly have very high dielectric constants. The first class are standard dielectrics, particularly barium titianate to which particles of metal have been added (8-10). It was empirically demonstrated that such mixed materials can have dielectric constants nearly ten times larger than the host ceramic (9).

Two contradictory models have been proposed to explain the impact of metal particle addition. The earlier model, which indicated this effect will only be observed for nano-scale metal inclusions (11-13), but, qualitatively, the same result was found later mixing 50 micron Ni particles with barium titanate (9). This led to the development of a different model, a ‘percolation’, model. Purportedly, some material properties, in particular dielectric constant, will diverge as the loading of the minority material (metal particles in this case) approaches the percolation limit of approximately 27% by volume (14-16). It is not clear that percolation (17) can explain the results of the present work as it is not clear what material in the NPS capacitors would be creating a percolation path, nor why that component of the mix was always near the percolation limit.

A second class of materials with high dielectric constants are 'colossal dielectric constant' materials, that is materials with dielectric constants greater than 1000. These include high Tc superconductors (18,19), some multi-metal oxides (20) and some nano-porous materials, such as metal organic framework materials (21). Dielectric constants higher than $10^5$ have been observed, but only over a very limited range of temperatures and frequencies, and only for single crystal materials prepared with special electrodes (22). The more accessible powder based colossal dielectric materials never appear to have dielectric values significantly greater than $10^4$. Although the potential use of these materials for electrical energy storage has been suggested (22), some of the data needed to evaluate this potential, such as maximum operating voltage, are not available.

It appears that for 'colossal' dielectrics there are two measurable dielectric values, intrinsic and extrinsic, and the 'intrinsic' values, rarely above 100, are the operative value for ~0 Hz, that is energy storage. Indeed, it is apparent that the dielectric constants/capacitance values are taken directly from commercial meters, which employ complex algorithms to derive their outputs, and generally reference very low voltages. Also, there is reason to believe that the dielectric 'constant' in some materials is not truly constant, and can be a sharp function of voltage. Indeed, in the present study the value of dielectric is clearly a function of voltage: very low above about 1 volt, very high between 1 volt and about 300 mV, and extremely high below 300 mV.

The origin of colossal dielectric constants is a matter of considerable discussion, and is only partially summarized here. One consensus is that the intrinsic dielectric properties of these materials are not 'colossal' and that the colossal values arise from a range of extrinsic properties such as the high dielectric values of interfaces, including interface between dielectric and electrode and grain boundaries (22,23). Other extrinsic high dielectric sources derive from the ability of trapped molecules, including water, to diffuse in porous networks in response to changing fields. Apparently, the process of production of some of these materials can increase the extrinsic dielectric value significantly (21). One model of the origin of the colossal dielectric constants is 'colossal polarization' (24). This model does seem to pertain to the observations herein, and in fact is virtually the same as that presented herein.

The above examples are of interest as they are empirically consistent with the finding of the present work that inhomogeneous materials can have higher dielectric constants than homogeneous materials. However; none of the particular models used to explain the data in the earlier works appear to readily apply to the data set presented in this paper. Indeed, there are no metal inclusions in the super dielectric NPS materials. The finding that the dielectric constant in the NPS material is independent of dielectric thickness, thus that capacitance increases with decreasing thickness, mitigating against the ceramic/electrode interface, which never changes, contributing significantly to super dielectric behavior. Observations, such as the impacts of drying and re-wetting on the measured dielectric constant are very difficult to explain with any model other than 'colossal polarizability'.

SDM are actually based on the empirical observation that the best practical single material dielectrics are the most polarizable, electrically insulating, materials. Uniquely, however, SDM are designed multi-material mixtures, not solid single crystal materials.

They are not even single phase, as they require both liquid and solid. One component, the ionic liquid, aqueous NaCl in this study, contains the 'polarizable element' in the form of mobile ions in a liquid solution. The other component, the highly porous insulating solid, is the physical framework, or skeleton, that holds the polarizable elements in place. Hence, SDM were *designed* to create the most polarizable, electrically insulating, multi-material.

The effort to study the impact of salt concentration on SD behavior is consistent with the model (Figure 7). In the figure, a schematic x-section of one of the SD created for this study, water drops of a variety of sizes and shapes are shown, each occupying a pore within the alumina framework. An enlargement of one drop shows the positive charged ions migrating toward the electrode with the negative field, and the negative ions migrating in the opposite direction. If the degree of polarization is key, then more ions should allow the formation of more strongly polarized drops. Indeed, a 'perfect' system would be one in which the permittivity of the drops approaches infinity, that is a perfect 'Faraday' response to an applied field.

It is important to note that there is no 'new theory paradigm in the model shown in Figure 7. The model of superdielectric behavior presented herein is essentially the classic model of dielectric behavior based on the polarization of the dielectric material (25,26). The only modification is one of 'particulars'. That is, it is the first time that the model has been applied to polarization of charges in ion-containing liquid drops contained in the pores of materials with decidedly low dielectric constants.

Note how the polarization model squares neatly with the basic definition of capacitance: The number of charges that are held on a given structure at a particular applied voltage, or inversely the number of charges required to create a particular potential on that structure. The super dielectrics in the NPS capacitors clearly require a dramatic increase in the number of charges, relative to a normal dielectric such as barium titanate, required to produce a particular field/potential, hence a dramatic increase in the capacitance.

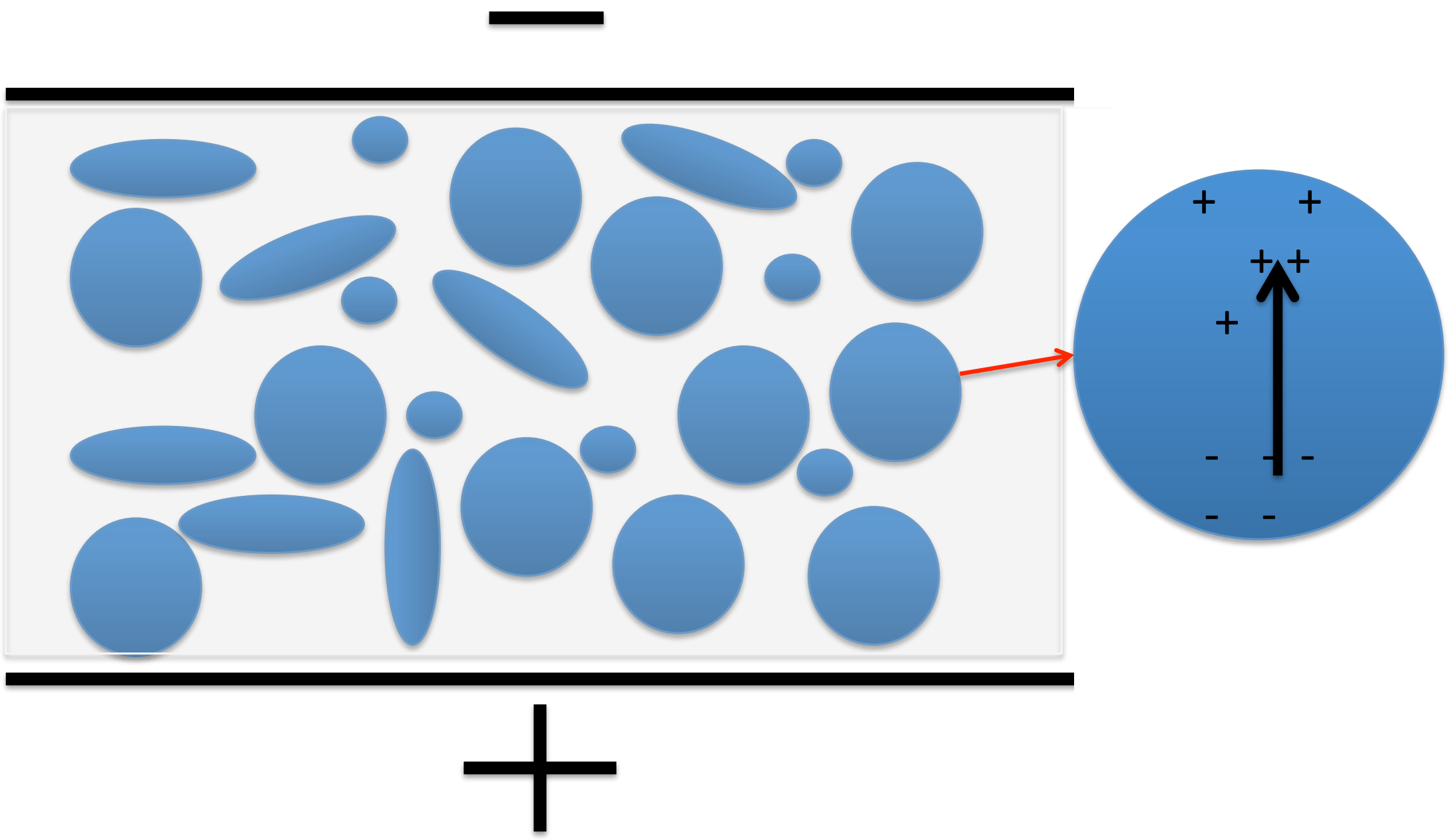

**Figure 7**- *X-Section Model of Super Dielectric in a Parallel Plate Capacitor*. In this model, alumina contains pores (circles, ellipses) are filled with an aqueous NaCl solutions. In an applied field, ions in each pore (right side expanded view) migrate to create a dipole that opposes the applied field.

In each water drop dipoles ‘oppose’ the applied field. The opposing fields result in a decrease in the net field. As more charges are added the dipoles grow larger, hence the net field grows slowly. Indeed, the experimental results collected in this study suggest that to create a specific potential between the metal electrodes it takes on the order of a 25 billion times more charges on the electrodes in an NPS capacitor containing polarizable drops than required in an air filled capacitor of the same dimensions. This suggests the ‘opposing’ field of the drops that mirrors the field created by charges on the electrode plates is very, very effective. In other words, the ‘permittivity’ of the drops containing dissolved ions is nearly infinite.

Additional data supports this model. First, is the finding that once all the water in the dielectric evaporates, the capacitance drops more than nine orders of magnitude. Moreover, simply re-adding water to the capacitor nearly completely restores the original capacitance. Consistent with the model, it is clear that once the media for charge transport, water, is removed, no ‘dipole mirror’ can form, hence there is no increase in capacitance. The net capacitance is probably that of alumina, very low. Second, the model does suggest that more ions in solution will strengthen the dipoles, making them more effective. It is clear that more salt does increase the dielectric value, but only to a point. In turn, this suggests the model presented is not sufficiently complex to explain all observations.

Another consistent observation: there is no significant capacitance above ~1.1 V. It is postulated in the model that this is caused by a breakdown of the water in the pores. It is well known that when breakdown occurs an electrolyte solution becomes highly conductive and can no longer sustain a voltage, or a dipole. Thus, the question becomes, is the observed maximum voltage consistent with the breakdown of an aqueous solution of NaCl? These three values enable a calculation: i) The breakdown voltage of distilled water is ~65 $*10^6$ V/m (27), ii) a solution of NaCl breaks down at a voltage about 10% lower than distilled water (28), iii) the average pore size in the alumina is approximately 250 Å, or 25 $*10^{-9}$ m. These values indicate that average drop should not be able to sustain a voltage greater than ~1.3 V. This is very close to the observed value.

Another observation is that there is a ‘break’ in the capacitance at approximately 300 mV. That is, below this value the capacitance increases significantly. One possible explanation for this observed behavior is that the alumina pore distribution is bimodal. One set of pores has an average diameter of perhaps 250 Å, and a second set an average diameter of 80 Å. The smaller pores do not become ‘effective’ in creating capacitance until the voltage drops below the breakdown voltage, which for 80 Å pores is about 300 mV. Or, perhaps all the drops are elliptical, with a long axis of 250 Å, and a short axis of 80 Å. Breakdown tends to occur across either the long or short axes depending on orientation relative to the applied field. In any event, the proposed model is reasonably close to observations, particularly given the uncertainty associated with measurements of pore size in powders.

The final point to consider is the potential energy density of an NPS capacitor. In order to compare the energy density of an NPS Capacitor to a battery or supercapacitor these values found in the literature of technology are needed: i) it is common to make electrostatic capacitors that are 0.5 micron thick using ceramic powder, ii) the other components of ceramic capacitor stacks (electrodes and insulators) are collectively of the same order of size or smaller than the ceramic layer. Next, from the standard equation for energy in a capacitor:

$E(\text{Joules}) = 0.5\ CV^2$

and the standard equation for parallel plate capacitance:

$C=\varepsilon\ \varepsilon o A/t$

the following key equation is derived:

Energy (J)/Volume ($m^3$)= $0.5\ \varepsilon\ \varepsilon o\ V^2/t^2$

Where ε is the dielectric constant, εo the permittivity of free space ($8.85\ 10^{-12}$ F/m), V is volts and t is the thickness of the layer in meters. Thus, given a conservative ultimate thickness of an NPS capacitor of 2.5 μ, assuming only 1 V potential and assuming that only one-half the volume of a ceramic capacitor stack is the dielectric layers, yields an astounding value, approximately ~9,000J/$cm^3$. Moreover, it is likely that higher voltages can be achieved by use of larger pore sizes, better electrolytes, etc. such that this energy density is unlikely to be anywhere near the ultimate value that can be attained with this technology. Indeed, if the theory presented herein is correct, an ultimate energy density north of 50,000 J/$cm^3$ is possible.

Often, for example in Ragone plots (29), the energy density of various devices, e.g. supercapacitors, Li Ion batteries, fuel cells, etc, are compared using the units Wh/kg. Assuming that the ultimate NPS capacitor stack, defined by the above conservative ultimate parameters of 1 V, and 2.5μ thickness is, by volume, 50% aluminum, 25% $H_2O$ and 25% alumina, yields an average density of ~4 g/$cm^3$. This yields >700 Wh/kg. Presently, commercial Li ion batteries deliver less than 500 Wh/kg. On a volumetric basis the contrast is even greater. The best Li Ion batteries are rated at ~1000 Wh/liter, whereas the NPS capacitor (1 V, 2.5μ thickness, etc.) is 2,500 Wh/liter. Hence, it is likely that the current generation is far from optimized and significant increases in energy density for NPS capacitors may occur in the near future.

CONCLUSION- The most significant finding of the present study is empirical: a pH neutral aqueous solution of NaCl in a porous alumina constitutes a super dielectric material. Moreover, the dielectric constant was a function of salt concentration, reaching a maximum value, at nearly one volt, of greater than 2.5 $10^{10}$ (25 *billion*), establishing that super dielectrics have dielectric constants orders of magnitude greater than any previously observed. This suggests that super dielectric materials, as per earlier suggestion, are a broad class of materials: porous, non-conductive materials saturated with ion-containing liquids. Moreover, they can be designed/optimized for energy storage by changes in the identity of the salt, concentration of the salt, the identity of the refractory oxide, the pore distribution in the refractory oxide, identity of the electrolyte, etc. All of these empirical results are consistent with a simple model of ‘colossal polarization’, in which the super dielectric behavior is explained to result from polarization of ions in the pore-filling liquid drops. The giant dipoles so formed ‘oppose’ the applied field, leading to a dramatic net increase in the amount of charge stored on the electrodes required to create a given net voltage.

The empirical findings also have practical consequences. Thin layer NPS capacitors will have energy densities several times greater than the best lithium ion batteries presently available, assuming the dielectric values obtained in the present work are sustainable at layer thicknesses (e.g 2.5 micron) widely found in today’s ceramic capacitors. This raises a question: Can batteries be replaced by more robust, less expensive, NPS capacitors?

References-